\documentclass[12pt,a4paper]{article}
\usepackage[utf8]{inputenc}
\usepackage[T1]{fontenc}
\usepackage{graphicx}	
\usepackage{amsmath}	
\usepackage{amssymb}
\usepackage{hyperref}
\usepackage{authblk}

\title{Natural Reference Frames within Video Analysis}

\author[1]{Fernando Saliby}
\affil[1]{Departamento de Ci\^encias da Natureza, Universidade Federal Fluminense, Rua Recife, Lotes 1-7, Jardim Bela Vista, Rio das Ostras - RJ, 28895-532, Brasil}
\affil[ ]{\texttt{fsimoni@id.uff.br}}

\begin{document}

\maketitle

\begin{abstract}
This study explores an alternative approach to video-based motion analysis using natural reference frames rather than relying on manual alignment. We demonstrate how the motion data itself can reveal optimal reference frames, connecting fundamental physical principles with data analysis. We present three classical mechanics experiments: uniform motion along a track, where the motion direction naturally defines the axis; circular motion of a bicycle wheel, where the center of rotation emerges as the reference origin; and projectile motion from a juggling tutorial, where gravity determines the vertical direction. This approach demonstrates how physical constraints and symmetries can naturally emerge from experimental data, providing a viable approach for introductory physics laboratories.
\end{abstract}

\section{Introduction}\label{sec:intro}
Video-based motion analysis in introductory physics laboratories often begins with manual reference frame alignment, where axes are positioned according to the system's apparent symmetries (e.g., matching expected motion directions or placing origins at visually identified points). While this approach leverages the system's physical properties, it relies on visual inspection. In physics, however, we can let the measured data itself reveal these same symmetries and its optimal reference frame. We call these ``natural reference frames'', as they emerge directly from the system's dynamics. This approach not only reduces systematic uncertainties but also offers students an opportunity to explore how fundamental physical principles manifest through data analysis.

In this study, we present three experiments commonly found in introductory mechanics laboratories where natural reference frames emerge from the measured motion data. For one-dimensional motion along a track, its direction defines the natural axis. In circular motion, the center of rotation sets the origin. And in projectile motion, the gravitational field determines the vertical direction. These simple experiments illustrate how identifying natural reference frames from data can be incorporated into introductory physics laboratories.

Video analysis measurements can be affected by various systematic effects. Geometric and optical distortions, such as perspective and parallax effects, and motion plane misalignments have been studied in detail~\cite{Martin2020,Stephens2019}. Digital imaging artifacts from modern CMOS cameras, such as rolling shutter and motion blur effects, can also influence the measurements~\cite{Rengarajan2014}. However, in introductory laboratories, a detailed treatment of these effects typically extends beyond the pedagogical objectives.

Since our goal here is to explore the use of natural reference frames in educational settings, we prioritize conceptual clarity over detailed error analysis. In our first two experiments (uniform and circular motion), we observe that while systematic effects are present, they do not obscure the relationship between reference frame choice and the system's behavior. Our third experiment analyzes projectile motion using 14 different throws extracted from a juggling tutorial video~\cite{juggling}, exploring natural reference frames in a real-world scenario while providing multiple measurements under similar conditions.

For the first two experiments, parameter uncertainties are obtained from least-squares fits, with the covariance matrix scaled to match the residuals' sample variance. This approach provides an estimate of statistical uncertainties, although it does not account for the systematic effects discussed above. For the projectile motion analysis, we report the standard deviation across the 14 throws as a measure of variability, reflecting the spread of repeated measurements under similar conditions.

All data were extracted using the Tracker software~\cite{Tracker} and are available, together with the analysis codes, on GitHub\footnote{\url{https://github.com/fsaliby/Natural_Reference_Frame_within_Video_Analysis}}. In the sections that follow, we present each experiment and discuss how natural reference frames can be identified from the data in introductory physics laboratories.

\section{Three Natural Frames}\label{sec:Three_Natural_Frames}

In this section we present three scenarios from introductory mechanics where natural reference frames emerge from the system's physical constraints: uniform, circular, and projectile motion.

\subsection{Uniform Motion}\label{sec:Uniform_Motion}

The first experiment uses a glider on an air track to study one-dimensional motion. The motion occurs in two phases: initially, the glider accelerates when pulled by a hanging weight; after the weight reaches the floor, the glider moves uniformly along the track. We focus our analysis on this second phase, where the constant velocity allows us to clearly demonstrate the role of reference frame selection. During 1.5 seconds of uniform motion, we collected 47 data points. The simplicity of this setup, where the track itself defines a clear axis for motion analysis, makes it particularly accessible for introductory physics courses, even for students with minimal background.

To determine the track's direction, we started with an arbitrary reference frame in Tracker and analyzed the data in two complementary ways. The first approach fits a straight line to the trajectory, \(y = \alpha x + \beta\), where the angle between the track and the original reference frame's horizontal axis is given by \(\theta = \arctan(\alpha)\). Note that this method only provides the orientation of the path, without indicating the actual direction of motion. As a consequence, the angle is constrained to the interval between \(-\pi/2\) and \(\pi/2\), corresponding to the fourth and first quadrants. This geometric approach can be applied to any type of one-dimensional motion, as it relies only on position coordinates.

The second approach uses the time evolution of position coordinates to determine both orientation and direction of motion. By fitting straight lines to \(x(t)\) and \(y(t)\) data, we obtain the velocity components \(v_x = (35.80 \pm 0.02)\) cm/s and \(v_y = (-15.64 \pm 0.01)\) cm/s. The angle between the track and the original reference frame's horizontal axis is obtained through \(\theta = \arctan(v_y/v_x)\), yielding $\theta = -23.60^{\circ} \pm 0.02^{\circ}$. This result, consistent with the first approach, shows that the glider's motion is in the fourth quadrant.

After determining the path orientation, we align our coordinate system with the natural reference frame by rotating it through the measured angle \(\theta\). First, we center the data by subtracting the initial position $x_0$ and the mean $\overline{y}$ value. Then, we apply the rotation transformation~\cite{Arfken2005}:
\begin{equation}\label{eq:rot_matrix}
R = \begin{pmatrix} 
\cos\theta & \sin\theta \\
-\sin\theta & \cos\theta
\end{pmatrix},
\end{equation}
where the coordinates in the rotated frame are given by $(x^{\prime}, y^{\prime}) = R(x - x_0, y - \overline{y})$.

\begin{figure}[t]
    \centering
    \includegraphics[width=0.8\textwidth]{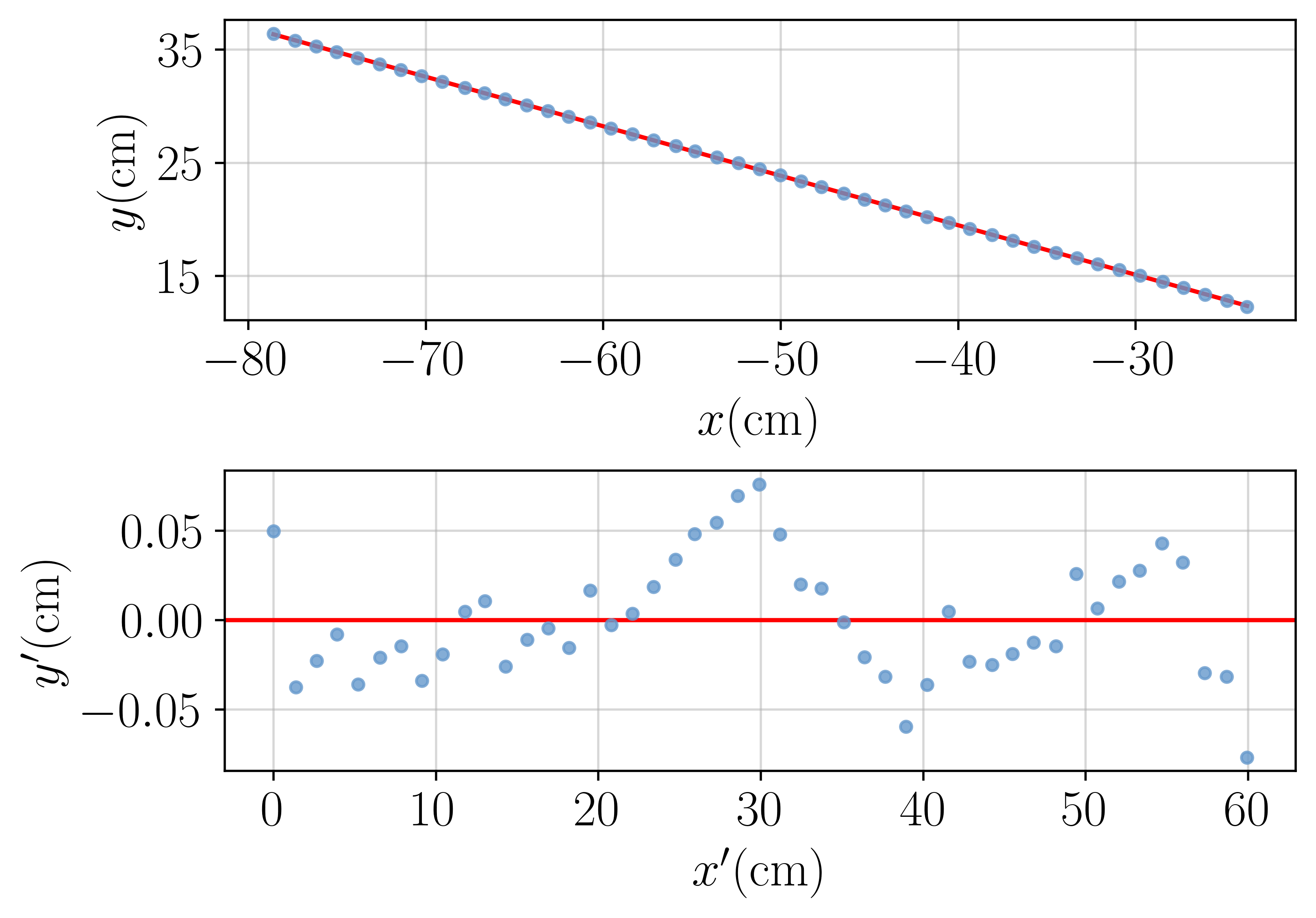}
    \caption{Uniform motion of the glider in two reference frames. Top: trajectory in the original arbitrary reference frame with the fitted line (red). Bottom: the same data after rotation to the natural reference frame, where $x^{\prime}$ is aligned with the track direction.}
    \label{fig:Uniform_Motion}
\end{figure}

Figure~\ref{fig:Uniform_Motion} shows the trajectory before and after the coordinate transformation. The upper panel shows the original data with its linear fit, illustrating the orientation between the motion and the arbitrary reference frame. The lower panel presents the same data in the rotated frame, where the motion occurs along $x^{\prime}$. The $y^{\prime}$ coordinates in the rotated frame correspond to the residuals of the linear fit shown in the upper panel, with a standard deviation of 0.03 cm from $y^{\prime}=0$.

In the original reference frame, the velocity magnitude is \(|v| = (39.07 \pm 0.02)\) cm/s. After the rotation, we obtain \(v_{x^{\prime}} = (39.07 \pm 0.02)\) cm/s and \(v_{y^{\prime}} = (0.00 \pm 0.01)\) cm/s. The vanishing \(v_{y^{\prime}}\) component and the compatibility between \(|v|\) and \(v_{x^{\prime}}\) indicate that we have successfully identified the natural reference frame where the motion occurs along a single direction.

\subsection{Circular Motion}\label{sec:Circular_Motion}

While the previous experiment explored natural reference frames in linear motion, our second experiment analyzes circular motion using a bicycle wheel, where the natural reference frame emerges from the center of rotation. We tracked the wheel's valve as it provided an unambiguous reference point for video analysis. During 2.6 seconds of motion, we collected 80 data points covering nearly one complete revolution, limited by the physical constraint of the bicycle fork. For the initial analysis, we positioned the Tracker's reference frame origin at the visually estimated center of rotation, following standard practice~\cite{Silva2018}.

\begin{figure}[t]
    \centering
    \includegraphics[width=0.8\textwidth]{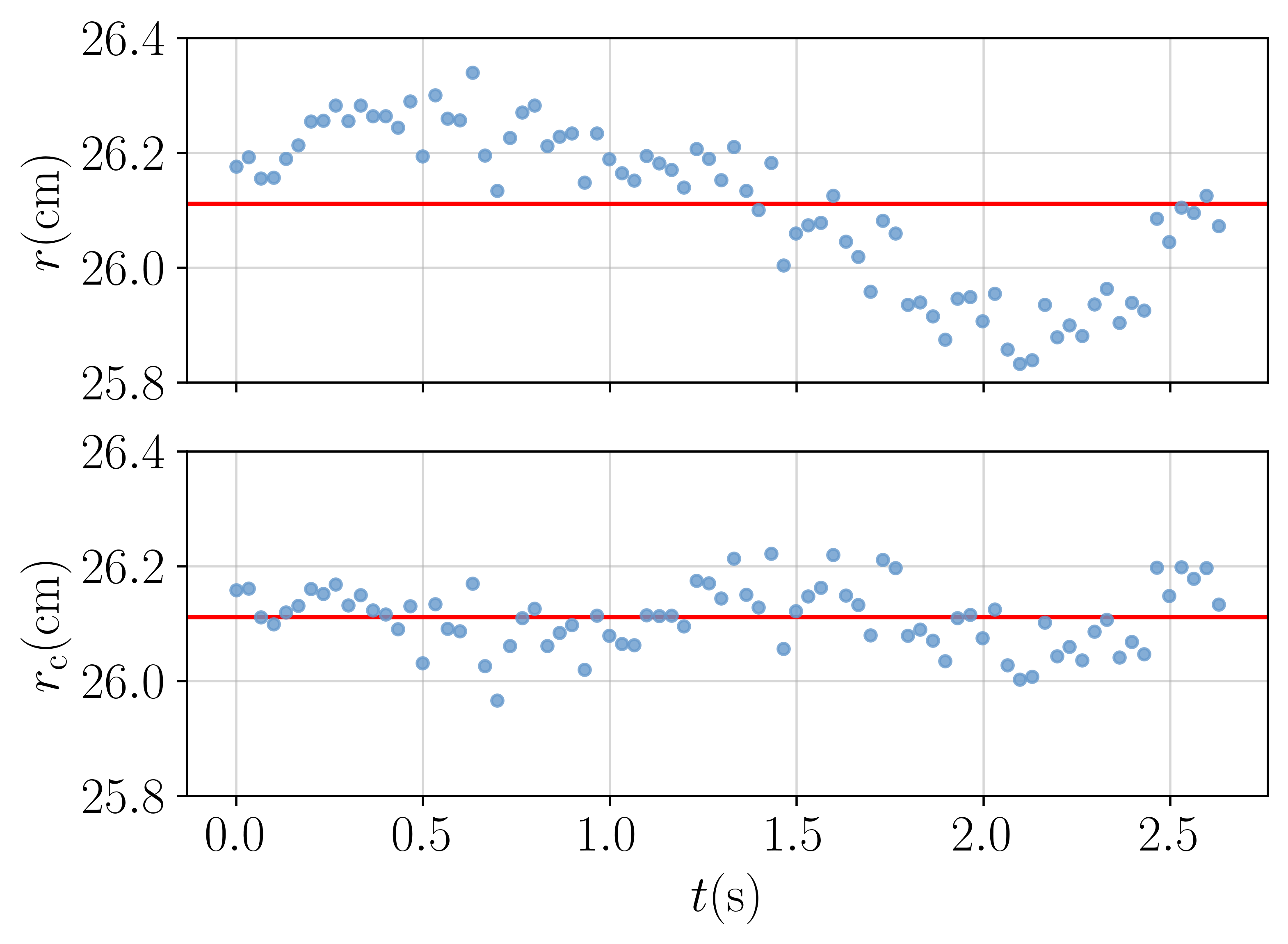}
    \caption{Time evolution of the distance magnitude for the valve. Top: Distance calculated from the original reference frame, showing oscillatory behavior due to the displacement from the true center of rotation. Bottom: Distance calculated after correcting the origin to the fitted center of rotation, reducing the standard deviation by a factor of 2.5. The horizontal red lines represent the mean radius in each reference frame.}
    \label{fig:radius_evolution}
\end{figure}

Figure~\ref{fig:radius_evolution} shows the time evolution of the distance magnitude from the origin, \(r(t)=\sqrt{x(t)^2 + y(t)^2}\), calculated from the original reference frame. This quantity exhibits oscillations with a frequency matching the wheel's rotation, suggesting a displacement between our chosen origin and the true center of rotation. The observed oscillation completes nearly one cycle, reflecting the wheel's motion.

For a displacement \((x_0,y_0)\) from the true center, the measured distance is given by
\begin{equation}\label{eq:displaced_radius}
    r(t) = \sqrt{x_0^2 + y_0^2 + R^2 + 2R \left[ x_0 \cos \theta(t) + y_0 \sin\theta(t) \right]},
\end{equation}
where $R$ is the true radius of the circular trajectory and $\theta(t)$ is the angular position. In this reference frame, the mean distance and its standard deviation are \((26.1 \pm 0.1)\) cm, with the distance magnitude varying by 0.5 cm throughout the motion.

To estimate the true center of rotation, we fitted a circle to the trajectory by minimizing the sum of squared distances from the points to the circle~\cite{Umbach2003}, assuming Gaussian-distributed residuals. These residuals are defined as
\begin{equation}\label{eq:circle_residuals}
    \varepsilon_i(x_0,y_0,R) = \sqrt{(x_i - x_0)^2 + (y_i - y_0)^2} - R.
\end{equation}
The fit indicates a displacement of \((x_0, y_0) = (-0.021 \pm 0.09, -0.169 \pm 0.009)\) cm from our initial reference frame, with a radius of \(R = (26.111 \pm 0.006)\) cm. The bottom plot of Figure~\ref{fig:radius_evolution} shows the distance magnitude after correcting for this displacement. In this natural reference frame, centered at the fitted rotation axis, the distance standard deviation decreases from 0.14 cm to 0.05 cm.

The circular fit reduces the distance magnitude fluctuations, but the bottom plot in Figure~\ref{fig:radius_evolution} shows a subtle systematic oscillation pattern around the mean radius. The original data exhibited one oscillation per revolution, while the centered data shows two oscillations per revolution. This behavior suggests that the trajectory might be better described by an ellipse rather than a circle, illustrating how reference frame choice can expose subtle features of the motion. An analysis of the trajectory's shape and its deviations from circularity is presented in Appendix~\ref{sec:app}.

\subsection{Projectile Motion}\label{sec:Projectile_Motion}

For our third experiment, we move from controlled conditions to a real-world scenario: analyzing projectile motion using 14 juggling ball throws extracted from a tutorial video~\cite{juggling}. In this case, gravity provides a natural definition of the vertical direction, illustrating another way physical constraints can guide reference frame selection. The analysis covered approximately 5 seconds of motion, with a total of 242 tracked points distributed among the throws.

Similarly to the circular motion experiment, we began with a manual alignment of the Tracker reference frame with the apparent vertical direction of the video. For calibration, we used the bialar breadth (the distance between the outer edges of the nose wings) as a scaling reference. This anthropometric measurement has a documented value of \((3.678 \pm 0.342)\) cm for North American adult males~\cite{young1993}.

\begin{figure}[t]
    \centering
    \includegraphics[width=0.8\textwidth]{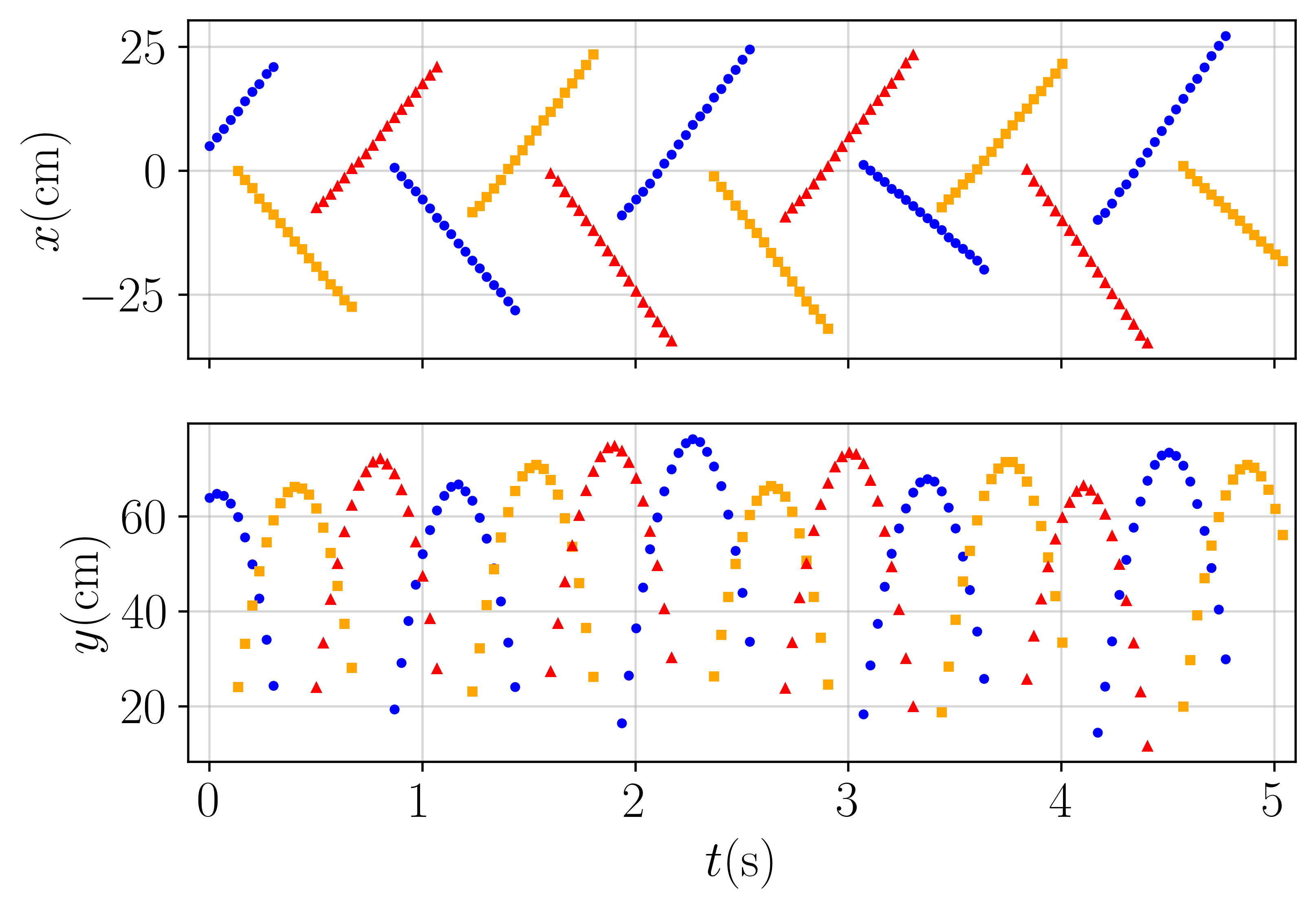}
    \caption{Time evolution of coordinates in the manually-aligned reference frame. Top: horizontal motion showing approximately linear trajectories. Bottom: vertical motion exhibiting the characteristic parabolic paths of projectile motion. Different markers and colors identify the three distinct balls used in the performance.}
    \label{fig:juggling}
\end{figure}

Figure~\ref{fig:juggling} shows the time evolution of coordinates for all 14 throws using three distinct juggling balls, each represented by different colors and markers. In our initial manually-aligned reference frame, the horizontal motion (top panel) exhibits approximately linear behavior, while the vertical motion (bottom panel) shows the characteristic parabolic trajectories of projectile motion. 

To determine the natural vertical direction independently for each throw, we applied a systematic approach based on the absence of horizontal acceleration in projectile motion. We analyzed rotated coordinate frames at different angles \(\theta\), measured counterclockwise from the positive \(x\)-axis of the initial reference frame. For each angle, the rotated horizontal coordinate is given by~\cite{Arfken2005}
\begin{equation}
x^{\prime}(\theta) = x\cos \theta + y\sin \theta .
\end{equation}

To find the natural horizontal direction, we analyze the motion in each rotated frame by fitting a quadratic function to \(x^{\prime}(t)\):
\begin{equation}
x^{\prime}(t) = \frac{1}{2}a_{x^{\prime}}t^2 + v_{x^{\prime}} t + x_0^{\prime},
\end{equation}
where the coefficient \(a_{x^{\prime}}\) represents the acceleration in this direction. The natural horizontal axis corresponds to the angle \(\theta_0\) where \(a_{x^{\prime}} = 0\). We then determine the vertical axis by requiring positive acceleration in the gravitational direction, completing our alignment with the natural reference frame.

Similarly to the Uniform Motion experiment (see Section~\ref{sec:Uniform_Motion}), the vertical direction could be determined by analyzing the spatial trajectory \(y(x)\) instead of the temporal evolution \(x(t)\) and \(y(t)\). We chose the temporal analysis as it employs kinematic equations commonly used in introductory physics courses, rather than the more complex conic section analysis.

After rotating to the natural reference frame, we fitted linear functions to the horizontal motion and quadratic functions to the vertical motion for all throws. The fits yield coefficients of determination \(R_{x^{\prime}}^{2} > 0.9994\) for the horizontal direction and \(R_{y^{\prime}}^{2} > 0.9978\) for the vertical direction. The fit residuals show standard deviations of 0.1 cm for horizontal motion and 0.4 cm for vertical motion. The larger vertical spread may be related to motion blur effects, as the initial vertical velocities are typically higher than horizontal velocities in these throws.

Our systematic analysis across all throws yields an angle of \(\theta_{0} = (0.2^{\circ} \pm 0.1^{\circ})\) between the manually-aligned and natural vertical directions, where the uncertainty represents the standard error of the mean across the 14 throws. This small angle indicates that our initial manual alignment was remarkably close to the natural vertical direction, consistent with recent studies on human perception of verticality. These studies show that humans can identify the true vertical direction with typical deviations of less than \(3^{\circ}\)~\cite{vanderWaal2024}.

The measured gravitational acceleration of \(\hat{g} = (11.40 \pm 0.08) \, \text{m}/\text{s}^{2}\) shows good consistency across the 14 throws. The standard deviation of \(0.3\) m/s\(^2\) in individual measurements indicates the juggler's consistency in throwing conditions throughout the performance. However, the measured value is 16\% higher than the expected \(9.8 \, \text{m}/\text{s}^{2}\). This systematic deviation can be attributed primarily to parallax effects, as the plane of ball motion differs from the plane where the scaling reference was measured, and potential differences between the juggler's bialar breadth and the average value used for calibration. 

The parallax correction follows the geometric relationship~\cite{Stephens2019}
\begin{equation}
g_{\text{true}} = \hat{g} \left( \frac{D}{D + d_o} \right),
\end{equation}
where \(D\) is the distance from the camera to the motion plane and \(d_o\) is the distance from the reference plane (the juggler's nose) to the motion plane. Assuming \(D = 2\) m and \(g_{\text{true}} = 9.8\) m/s\(^2\), we obtain \(d_o \approx 30\) cm, comparable to typical juggling distances. The documented 9.3\% variability in bialar breadth~\cite{young1993} would lead to \(g\) estimates between 10.3 and 12.4 m/s\(^2\). These calculations suggest that both parallax and scaling uncertainties contribute to the observed deviation in our \(g\) measurement.

While both systematic effects (parallax and scaling) alter the measured value of \(g\), they do not affect the determination of the vertical direction \(\theta_{0}\), preserving the main result of our analysis.

\section{Conclusion}\label{sec:Conclusion}

The three experiments presented in this study explore how motion data can guide the identification of natural reference frames in video-based analysis, providing an alternative to manual alignment. Each experiment examines a different aspect of this data-driven approach: in uniform motion, the trajectory indicates the natural axis; for circular motion, the center of rotation emerges from the system's geometry; and in projectile motion, gravity defines the vertical direction. This approach offers opportunities to explore physical aspects of the system while working with fundamental principles of motion and symmetry.

In uniform motion (see Section~\ref{sec:Uniform_Motion}), we applied two complementary methods to determine the natural reference frame: trajectory analysis and velocity components. These approaches provide independent estimates of the same physical quantity---the direction of motion---using concepts familiar to introductory physics students. The combination of these methods allows exploration of connections between kinematic and geometric descriptions of motion. Additionally, as shown in the bottom plot of Figure~\ref{fig:Uniform_Motion}, the aligned reference frame exhibits measurement residuals with a standard deviation of 0.03 cm around zero. These residuals provide an opportunity to discuss measurement uncertainty and the inherent variability in experimental data.

For circular motion (see Section~\ref{sec:Circular_Motion}), the initial manual alignment produced oscillatory patterns in the radial distance measurements, which indicated a displacement from the center of rotation. Through systematic circle fitting, we identified this displacement and reduced radial distance fluctuations from 0.14 to 0.05 cm. This reduction in fluctuations illustrates how data-driven reference frame selection can contribute to measurement precision.

In the projectile motion experiment (see Section~\ref{sec:Projectile_Motion}), the gravitational field provided a natural vertical reference. Our systematic determination of this direction, based on the absence of horizontal acceleration, yielded an angular deviation of \(\theta_{0} = (0.2 \pm 0.1)^{\circ}\) from manual alignment. While systematic effects affected the measurement of gravitational acceleration, they did not influence the determination of the vertical direction, suggesting that the identification of natural reference frames can be reliable even under non-ideal conditions.

The methodology presented here offers an alternative approach for introductory physics laboratories, combining data analysis with fundamental physics principles. Working with natural reference frames allows students to explore how physical constraints and symmetries manifest in experimental measurements, connecting theoretical concepts with practical data analysis.

\appendix
\section{From Circular to Elliptical}\label{sec:app}

When recording circular motion with a video camera, the quality of the measurements depends on the alignment between the motion plane and the camera sensor plane. A misalignment between these planes causes the circular trajectory to appear as an ellipse in the recorded images. In Section~\ref{sec:Circular_Motion}, after determining the center of rotation, we observed a systematic oscillation in the radius (see the bottom plot of Figure~\ref{fig:radius_evolution}). This appendix examines this effect by comparing circular and elliptical fits to the data.

The fitting procedure for the elliptical trajectory followed a methodology similar to that used in the circular fit, but required five adjustable parameters to fully characterize the geometric distortion: the center coordinates ($x_0$, $y_0$), the semi-major and semi-minor axes ($a$, $b$), and the ellipse orientation angle ($\theta_{\text{e}}$). These parameters provide a complete description of how a circular trajectory appears when viewed from an angle.

The elliptical fit to our data yields a center at $(-0.013 \pm 0.008, -0.169 \pm 0.007)$ cm, with semi-major and semi-minor axes of \((26.162 \pm 0.009)\) cm and \((26.067 \pm 0.009)\) cm, respectively, and an orientation angle of $\theta_{\text{e}} = -19^{\circ} \pm 4^{\circ}$. The residuals standard deviation decreases from 0.055 cm in the circular fit to 0.003 cm with the elliptical model, suggesting that perspective effects introduce measurable deviations from circularity.

The inclination angle between the wheel's motion plane and the camera plane can be determined from the ratio of the semi-minor to semi-major axes through the relation
\begin{equation}
\phi = \arccos \left( \frac{b}{a} \right),
\end{equation}
yielding $\phi = 4.9^{\circ} \pm 0.2^{\circ}$. This misalignment angle is consistent with the systematic deviations observed in the circular fit, illustrating how perspective effects can be quantified through geometric analysis of video measurements.

\section*{Acknowledgements}

The author acknowledges the assistance of Claude AI (Anthropic) in manuscript preparation, code development and brainstorming discussions.

\end{document}